# Novel Logical Method for Security Analysis of Electronic Payment Protocols

Yi Liu [a], Xingtong Liu [a], Lei Zhang [a], Jian Wang[a] and Chaojing Tang [a]

[a] College of Electronic Science and Engineering, National University of Defense Technology, No.47, Yanwachi street, 410073 Changsha, Hunan, China

**Abstract**

Electronic payment protocols play a vital role in electronic commerce security, which is essential for secure operation of electronic commerce activities. Formal method is an effective way to verify the security of protocols. But current formal method lacks the description and analysis of timeliness in electronic payment protocols. In order to improve analysis ability, a novel approach to analyze security properties such as accountability, fairness and timeliness in electronic payment protocols is proposed in this paper. This approach extends an existing logical method by adding a concise time expression and analysis method. It enables to describe the event time, and extends the time characteristics of logical inference rules. We analyzed the Netbill protocol with the new approach and found that the fairness of the protocol is not satisfied, due to timeliness problem. The result illustrates the new approach is able to analyze the key properties of electronic payment protocols. Furthermore, the new approach can be introduced to analyze other time properties of cryptographic protocols.

*Keywords:* electronic payment protocol; formal analysis; accountability; fairness; timeliness; logic reasoning;

# 1. Introduction

In recent years, an explosion of services provided over the Internet has a great importance of human daily life. These services are increasingly transferring customers' private and financial information over the network. A novel proof methodology to verify secure routing protocols has been proposed by Chen (Chen et al. 2015). Another category of network protocols to protect legitimate interests between traders also need to be verified with additional security properties. Electronic payment protocols provide technical assurance for security of electronic commerce. Sensitive information such as credit card numbers and password depends on the security of electronic payment protocols. The analysis and research on the security of electronic payment protocols has become an important issue in the field of information security (Patrick et al. 2016).

Compared with other security protocols, accountability, fairness and timeliness are additional security properties in electronic payment protocols. Accountability can provide sufficient evidence to resolve possible future disputes after the execution of protocol. It means that all parties cannot repudiate what they have done. Fairness ensures that no one can gain an advantage over other participants by misbehaving, which means either both the participants receive what they expect or nothing. Timeliness provides constraint of intervals during every step in protocol regulation to avoid the time difference utilizing by attackers.

Formal analysis is an effective method to verify electronic payment protocols thanks to its strict and effective characteristics. But current formal methods for analysis of electronic payment protocols lack the description and analysis of timeliness. Our approach focuses on the description and analysis of three security properties above. We add concise time expression and analysis method to existing logical method. The logic reasoning part in the process of the objective proof is based on Qin-Zhou logic method (Qing 2005; Zhou et al. 2001), and the time calculus part utilizes the method of algebra and set theory. The logical method and the algebraic method are two independent parts. They will not interfere with each other or undermine the correctness of the original method. The Netbill protocol is analysed with the our method, and the result shows that the protocol does not satisfy fairness due to timeliness problem. Then we elaborated that the defect can be fixed with careful specification of event time and waiting time.

The rest of this paper is organized as follows. Section 2 introduces related work in this area. Section 3 describes the concepts and definitions of the novel logical method. The logic analysis procedure is introduced in Section 4. The analysis process of the Netbill protocol is illustrated in Section 5. Section 6 concludes this paper and outlines our future studies.

**2. Related work**

Formal methods have already been used for security analysis of electronic payment protocols for decades (Kailar 1996). They can be divided into three categories as logic reasoning, model checking and theorem proving.

**2.1 Logic reasoning**

Logic reasoning is an important formal analysis method of electronic payment protocols up to present. Kailar logic (Kailar 1996) is the first analysis method designed for electronic payment protocols, which is mainly used to analyse accountability. But it ignored fairness in electronic payment protocols. Volker extended Autlog logic to be able to analyse accountability (Volker and Heike 1998). The famous Payword and SET protocol are analysed as examples. Qing-Zhou logic was proposed for the analysis of accountability and fairness together (Qing 2005; Zhou et al. 2001). Li added the time factor in SVO logic to make it able to analyse the timeliness of protocols (Li and Luo 2006). Wen put forward a modeling and analysis method of electronic payment protocols based on game logic (Wen et al. 2007). Chen combined logic reasoning with strand space model, introducing a new logic analysis method of electronic payment protocols (Chen 2010). A method applying Kailar logic in compositional analysis is presented by Gao for analysing the accountability and fairness of electronic payment protocols (Gao et al. 2013).

**2.2 Model checking**

The characteristic of model checking is easy to manipulate. Kremer apply model-checker MOCHA which supports the alternating transition systems and the alternating temporal logic to analyse accountability (Steve 2004). Xie utilized finite automaton to analyse ISI protocol and IBS protocol (Xie and Zhang 2004). Guo combined communication finite state machine with some new logic rules based on Qing-Zhou logic to analyse security properties of electronic payment protocols (Guo et al. 2010). Liu proposed an extended deterministic finite automaton which can also analyse security properties such as accountability and fairness (Liu et al. 2013). Nevertheless, due to the state space of the model checking method is limited, even if no attack method has been found, it does not mean the correctness of protocols.

**2.3 Theorem proving**

Theorem proving is regarded as an accurate method for security of cryptographic protocols. Papa integrated logic with process calculi for analysing electronic payment protocols (Papa et al. 2001). Chun used Coloured Petri Nets to analyze the Internet Open Trading Protocol (Chun and Jonathan 2004). Bella analysed the purchase protocol of SET with Isabelle and the inductive method (Giampaolo et al. 2006). Guttman applied strand space method to analyse the fairness of fair exchange protocols (Guttman 2012). Guo proposed a technique for modelling and verifying fair-change electronic payment protocols (Guo et al. 2009). On the other hand, theorem proving method is complicated and difficult to verify complex protocols.

## 3. Concepts and definitions

The definitions and symbols used in the new approach are denoted as follows:

Table 1. Basic symbols

| | | | |
|---|---|---|---|
| $A, B$ | Parties participate in protocol. | $\{m\}_K$ | Ciphertext of message $m$ encrypted with secret key $K$. |
| $TTP$ | Trusted third party. | $\tilde{K}$ | Dual key of $K$. |
| $m$ | Message transferred in protocol. | $(m, n)$ | Message $m$ is combined with message $n$. |
| $K_a$ | The public key of party $A$, which is used to verify the digital signature of $A$. | $K_a^{-1}$ | The private key corresponding to $K_a$. |
| $EOO$ | The non-repudiation evidence that is provided to the receiver in electronic payment protocols, which is used to prove that the sender has sent the message. | $EOR$ | The non-repudiation evidence that is provided to the sender in electronic payment protocols, which is used to prove that the receiver has received the message. |
| $T$ | Time of event. | | |

### 3.1 Time system

We denote the time when event occurs by adding a condition in the formula language, like $A \rightarrow m$ at $T$. $T$ is a time expression. This definition introduced the description of the occurrence time of sending and receiving message.

The time expression is defined as follows:

1. $x$ stands for a constant time element.

2. $X$ stands for a variable time element.

3. $X|TS$ means a time binding expression, while $TS$ is the scope of $X$.

4. $[T]$ is time expression, while $T$ is a time binding expression.

The constant time element is represented by a lower case *t*, and the variant time element is represented by a capital letter *T*. Time binding expression is a variable time element *X* with a certain value of constant time element as $t(t \in TS)$. In logical formula, the time expression [*X*|*I*] can be abbreviated to [*X*], and [*X*|{*x*}] can be abbreviated to [*x*], where *x* is a constant time element or a variable time element with bound value. The value of a variable time element is bound to the first operation in its formula.

**3.2 Protocol and environment**

*TTP*(Trusted third party) is a special party, which is regarded as a fair trusted third party. The bank or the arbitration organization can act as *TTP*. In general, we assume that all parties are dishonest except for *TTP*. They may interrupt the execution of protocols arbitrarily.

Communication channel is either reliable or unreliable, depending on the specific operating environment. Usually, the communication channel between general parties is unreliable, while it between the *TTP* and other parties is recoverable which means the message will be transmitted eventually.

Protocol statement defines what message should be sent and received by parties in the current round as follows:

$A \rightarrow B : m$ at *T* means *A* sent message *m* to *B* at *T*.

**3.3 Possession set**

$O_a$ stands for the possession set of party *A* participating in protocol. Assuming the protocol begins from $T_0$, the initial possession set of *A* is $O_a(T_0)$. When protocol runs to $T_x$, the possession set of *A* becomes $O_a(T_x)$. Besides, we define $O_a(T_e)$ is the final possession set of *A* at the end of protocol. The possession set of *A* contains the information inherited from last step and the message which is received or generated at present. It varies consecutively with the execution of protocol until $O_a = O_a(T_e)$.

The possession set of A changes from $O_a(T_y)$ to $O_a(T_x) \cap (T_y \leq T_x)$, which means $T_y$ is the moment before $T_x$. It follows the following rules:

(1) When the execution of protocol statement is $A \to B : m$ at $T_x$, $O_a(T_x) = O_a(T_y) \cup \{m\}$ if m is a new message generated by A, which means $m \notin O_a(T_y)$. Otherwise, $O_a(T_x) = O_a(T_y)$ if m is not a new message generated by A and $m \in O_a(T_y)$.

(2) When the execution of protocol statement is $B \to A : m$ at $T_x$ while $m \notin O_a(T_y)$, $O_a(T_x) = O_a(T_y) \cup \{m\}$. Otherwise, $O_a(T_x) = O_a(T_y)$.

## 4. Logic analysis method

### 4.1 Logic component

Our method consists of the following five logic components:

(1) $A \succ x$. A can make others believe in formula x by performing a series of operations without leaking any secret;

(2) $A \to m$ at T. A sends message m at T. The following implication is established in the process of analysis:

$$A \to (m,n) \text{ at } T => A \to m \text{ at } T \quad (1)$$

It means, A sends message m at T while A sends message (m, n) at T.

(3) $A \ni m$. A possesses message m.

(4) $A \leftarrow m$ at T. A received message m at T. The following implication is established as the second component:

$$A \leftarrow (m,n) \text{ at } T => A \leftarrow m \text{ at } T \quad (2)$$

(5) $\xrightarrow{K_a} A$. $K_a$ is the public key of A, which is used to verify the message signed by its private key $K_a^{-1}$.

### 4.2 Axiom system

The axiom system consists of 1 inference rule and 6 axioms. The inference rule is as follows:

$$(\vdash\varphi) \cap (\vdash(\varphi \Rightarrow \psi)) \Rightarrow \vdash\psi \qquad (3)$$

It illustrates $\vdash\psi$ can be obtained from $\vdash\varphi$ and $\vdash(\varphi \Rightarrow \psi)$. $\Gamma \vdash \psi$ represents $\psi$ can be deduced from the formula sets $\Gamma$. $\vdash\varphi$ indicates $\varphi$ is a theorem, which means $\varphi$ is established all the time. The inference rule above indicates that $\psi$ is theorem if $\varphi$ is theorem and $\varphi$ contains $\psi$.

The 6 axioms are as follows:

A1. $A \succ x \cap A \succ y \Rightarrow A \succ (x \wedge y)$

A2. $A \succ x \cap (x \Rightarrow y) \Rightarrow A \succ y$

A3. $A \ni \{m\}_{K_b^{-1}}$ at $T_x \cap A \succ \xrightarrow{K_b} B$ at $T_x \Rightarrow A \succ B \to m$ at $[T_Y | T_Y \leq T_X]$

A4. $A \succ B \to \{m\}_k$ at $T_x \cap A \succ B \to k$ at $T_Y \Rightarrow A \succ B \to m$ at $\max(T_X, T_Y)$

A5. $A \leftarrow m$ at $T \Rightarrow A \ni m$ at $T$

A6. $A \leftarrow \{m\}_K$ at $T \cap A \ni \tilde{K} \Rightarrow A \leftarrow m$ at $T$

The proof procedure of protocol properties is divided into two parts. The first part is called logical reasoning and the second part is called time calculus. The purpose of this procedure is to prove that the result obtained from logic reasoning satisfies the time constraints specified in time calculus.

### 4.3 Protocol analysis procedure

Protocol analysis consists of 5 steps as shown in Figure 1.

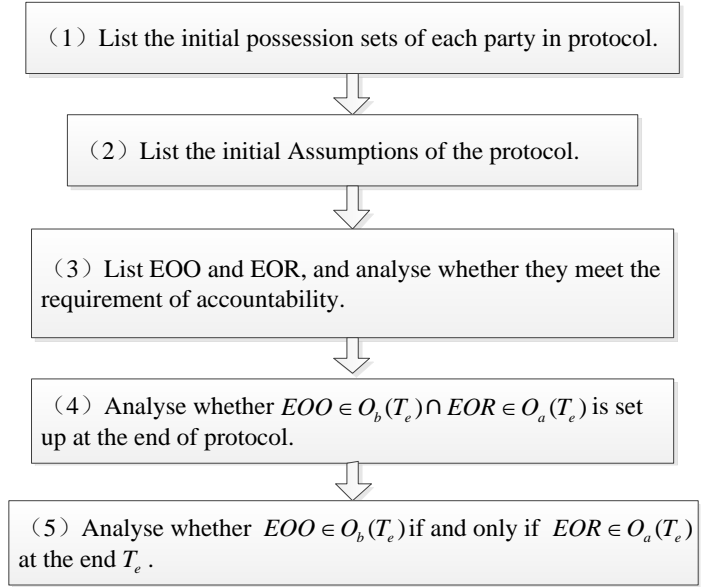

Figure 1. Procedure of protocol analysis

## 5. The Netbill Protocol analysis

The Netbill protocol is an electronic payment protocol proposed by Professor J.D.Tygar in Carnegie Mellon University for trading digital goods, including three participants: customer, merchant and the Netbill server (Sirbu and Tygar 1995). Its main steps are as follows:

(1) $C \rightarrow M : T_{CM}(C), \{PRD, TID\}_{K_{CM}}$ at $T_1$

(2) $M \rightarrow C : \{ProductID, Price, TID\}_{K_{CM}}$ at $T_2$

(3) $C \rightarrow M : T_{CM}(C), \{TID\}_{K_{CM}}$ at $T_3$

(4) $M \rightarrow C : \{Goods\}_k, \{h(\{Goods\}_k), EPOID\}_{K_{CM}}$ at $T_4$

(5) $C \rightarrow M : T_{CM}(C), \{\{EPO\}_{K_C^{-1}}\}_{K_{CM}}$ at $T_5$

(6) $M \rightarrow N : T_{MN}(M), \{\{\{EPO\}_{K_C^{-1}}, MAcct, k\}_{K_M^{-1}}\}_{K_{MN}}$ at $T_6$

(7) $N \rightarrow M : \{\{Receipt\}_{K_N^{-1}}, \{EPOID, CAcct\}_{K_{CN}}\}_{K_{MN}}$ at $T_7$

(8) $M \rightarrow C : \{\{Receipt\}_{K_N^{-1}}, \{EPOID, CAcct\}_{K_{CN}}\}_{K_{CM}}$ at $T_8$

$C$, $M$ and $N$ represent the customer, the merchant and the Netbill server respectively. $T_{CM}(C)$ is used to prove $C$ to $M$, and to establish a shared session key $K_{CM}$. The function of $T_{MN}(M)$ is similar to $T_{CM}(C)$. *PRD*(Product Request Data) is product request data. *TID* is transaction's ID and *ProductID* is product's ID. *Price* stands for the price the merchant required. *Goods* is the specific content of transmitted goods. *EPO* is a electronic purchase order, the plain part of which comprises customer identification *identity*, *ProductID*, *Price*, *M*, $h(\{Goods\}_k)$ and *h(PRD)*. *h(m)* stands for the hash value of *m*. The encryption part includes a payment instruction which can only be read by the Netbill server such as the customer account. *EPOID* is electronic payment ID, which is the globally unique identifier and will be used to uniquely identify the transaction in the database of Netbill. *CAcct* and *MAcct* stand for the customer and merchant's account respectively. *Receipt* includes the *Result* whether to accept this payment or not, which is returned from the Netbill server.

The analysis procedure of the Netbill protocol is detailed in the next subsection.

### 5.1 List the initial possession sets.

At the initial time of protocol operation, the initial states of *C* and *M* are

$$O_C(T_0) = \{K_C^{-1}, K_C, K_M, K_N, K_{CM}, K_{CN}\}$$
$$O_M(T_0) = \{K_M^{-1}, K_M, K_C, K_N, K_{CM}, K_{MN}\}$$
$$C \succ (\xrightarrow{K_M} M, \xrightarrow{K_N} N, C \xleftrightarrow{K_{CM}} M, C \xleftrightarrow{K_{CN}} N)$$
$$M \succ (\xrightarrow{K_C} C, \xrightarrow{K_N} N, C \xleftrightarrow{K_{CM}} M, M \xleftrightarrow{K_{MN}} N)$$

### 5.2 List the credible assumptions

T1: $A \succ N \to k \Rightarrow A \succ B \to k$

Assume that the Netbill server is fully in accordance with the protocol regulation and will not do anything harmful to any party. If *A* can prove that *N* has sent *k* to him, he can prove that the other party *B* has sent *k*.

T2: $A \succ B \rightarrow h(m) \Rightarrow A \succ B \rightarrow m$

According to the protocol, the sender sends $h(m)$ for the checksum of message $m$. The sender is able to calculate the checksum only when the sender owns the message $m$. So if $A$ can prove that $B$ has sent $h(m)$, then $A$ can prove that $B$ has sent message $m$.

## 5.3 List *EOO* and *EOR*:

$$EOO = (\{h(\{Goods\}_k)\}_{K_{CM}}, \{k\}_{K_N^{-1}})$$

$$EOR = (\{h(\{Goods\}_k)\}_{K_C^{-1}}, \{k\}_{K_N^{-1}})$$

Assume that $EOO \in O_C(T_e)$ is established at the end of protocol. Then $(\{h(\{Goods\}_k)\}_{K_{CM}}, \{k\}_{K_N^{-1}}) \in O_C(T_e)$ is satisfied, which means $C \ni \{h(\{Goods\}_k)\}_{K_{CM}}$ at $T_e$ and $C \ni \{k\}_{K_N^{-1}}$ at $T_e$.

Since $C \ni \{h(\{Goods\}_k)\}_{K_{CM}}$ at $T_e$, $C \succ C \xleftrightarrow{K_{CM}} M$ and axiom A3, then $C \succ M \rightarrow h(\{Goods\}_k)$ at $[T_\alpha | T_\alpha \leq T_e]$. According to T2, we obtain

$$C \succ M \rightarrow \{Goods\}_k \text{ at } [T_\alpha | T_\alpha \leq T_e] \tag{4}$$

Since $C \ni \{k\}_{K_N^{-1}}$ at $T_e$, $C \succ \xrightarrow{K_N} N$ and axiom A3, then $C \succ N \rightarrow k$ at $[T_\beta | T_\beta \leq T_e]$. According to the credible assumption T1, we obtain

$$C \succ M \rightarrow k \text{ at } [T_\beta | T_\beta \leq T_e] \tag{5}$$

Due to formula (4), (5) and axiom A4, we will get

$$C \succ M \rightarrow Goods \text{ at } \max(T_\alpha, T_\beta) \cap [T_\alpha | T_\alpha \leq T_e] \cap [T_\beta | T_\beta \leq T_e] \tag{6}$$

Assume that $EOR \in O_M(T_e)$ is satisfied at the end of the protocol, which means $M \ni \{h(\{Goods\}_k)\}_{K_C^{-1}}$ at $T_e$ and $M \ni \{k\}_{K_N^{-1}}$ at $T_e$ are satisfied. Then according to $M \succ \xrightarrow{K_N} N$, axiom A3 and credible assumption T1, we obtain

$$M \succ C \ni k \text{ at } [T_\gamma | T_\gamma \leq T_e] \tag{7}$$

Since $M \succ \xrightarrow{K_C} C$, axiom A3 and credible assumption T2, we will get $M \succ C \ni \{Goods\}_k$ at $[T_\theta | T_\theta \leq T_e]$. Due to formula (7) and axiom A6, we will get

$$M \succ C \ni Goods \text{ at } \max(T_\gamma, T_\theta) \cap [T_\gamma | T_\gamma \leq T_e] \cap [T_\theta | T_\theta \leq T_e] \tag{8}$$

Therefore, the design of *EOO* and *EOR* in the Netbill protocol can meet the requirement of accountability.

### 5.4 Analysis of accountability

Verify whether *C* and *M* can obtain the appropriate evidence at the end of protocol. After the fourth step of the protocol,

$O_C(T_4) = O_C(T_3) \cup \{h(\{Goods\}_k)\}_{K_{CM}} \cap [T_4 | T_4 \leq T_e]$, then $\{h(\{Goods\}_k)\}_{K_{CM}} \in O_C(T_e)$.

When the last step of the protocol is finished,

$O_C(T_8) = O_C(T_7) \cup \{\{Receipt\}_{K_N^{-1}}\}_{K_{CM}} \cap [T_8 | T_8 \leq T_e]$. Because of $C \ni K_{CM}$, we obtain $\{Receipt\}_{K_N^{-1}} \in O_C(T_e)$, and $\{k\}_{K_N^{-1}} \in O_C(T_e)$. Then $EOO \in O_C(T_e)$ is satisfied.

Similarly, according to the fifth step of the protocol, we will get

$\{h(\{Goods\}_k)\}_{K_C^{-1}} \ni O_M(T_e)$. And $\{k\}_{K_N^{-1}} \in O_M(T_e)$ will be obtained after the seventh step. Then we will get $EOR \in O_M(T_e)$.

Therefore, $EOO \in O_C(T_e) \cap EOR \in O_M(T_e)$ is satisfied when the protocol finishes.

### (5) Analysis of fairness

The fairness objective is:

$$EOO \in O_C(T_k) \text{ if and only if } EOR \in O_M(T_k)$$

All parties in protocol will wait for the next step after the last one. If there is no response, the protocol will terminate after a certain period of time *t* and clear the protocol records before. For achieving fairness, the following conditions have to be satisfied:

$$M \to \{\{k\}_{K_M^{-1}}\}_{K_{MN}} \text{ at } T_x \cap M \leftarrow \{k\}_{K_N^{-1}} \text{ at } T_y \cap (T_x \leq T_y \leq T_x + t_M) \quad (9)$$

$$C \to \{h(\{Goods\}_k)\}_{K_C^{-1}} \text{ at } T_x \cap C \leftarrow \{k\}_{K_N^{-1}} \text{ at } T_y \cap (T_x \leq T_y \leq T_x + t_C) \quad (10)$$

$t_M$ is the waiting time after the $M$ executes the 6th step of the protocol. $t_C$ is the waiting time after $C$ executes the 5th step. Since $N$ is in full accordance with the regulation of protocol, formula (9) must be established.

In formula (10), $T_x = T_5$, $T_y = T_7 = T_5 + t_5 + t_6$, $t_5$ and $t_6$ are the time delay after the 5th step and the 6th step. So we must make $t_5 + t_6 \leq t_C$ to make formula (10) established. But there is no restrict about the relationship among $t_5, t_6$ and $t_C$. There is a possibility to make $t_5 + t_6 > t_C$ no matter what the constant $t_C$ is specified. For example, if $C$ performs the 5th step in accordance with the regulation, $M$ could send $\{\{EPO\}_{K_C^{-1}}\}_{K_{MN}}$ to $N$ after it is timeout and acquire the evidence to prove $C$ has received the product $Goods$. But $C$ has deleted $\{Goods\}_k$ because of timeout. Even though $C$ has received $Receipt$ and the key $k$, he couldn't decrypt the ciphertext to obtain the product $Goods$.

Then the fomula (10) can't be established, which means the protocol can't achieve the fairness objective. The main reason is that the implementation of the protocol does not have specific constraints on the relevant event time in the process. In order to make up for the defect, we should carefully regulate the event time and the waiting time of the execution of protocol.

**5 Conclusion**

The analysis result of Netbill protocol show that the protocol does not satisfy fairness because of timeliness problem. The process illustrates how the new approach is applied to analyse the temporal relation between events in electronic payment protocols. It is not a simple logic method, but an integrated approach. The new approach is appropriate to guide the design of electronic payment protocols and fix the defects of original protocols.

The next step of our research is to analyse other electronic payment protocols with our method which are widely used in electronic commerce. At the same time, we will further study automated analysis tools to make it convenient for design and analysis of electronic payment protocols.

**Acknowledgements**

This work was sponsored by the National Natural Science Foundation of China (Project No. 61601476).